\documentclass{revtex4}

\usepackage{graphicx}
\begin{document}
% You should use BibTeX and revtex.bst for references
\bibliographystyle{revtex}

% number  and your conference paper identification number on the
% title page in preprint mode. Multiple \preprint commands are allowed.
%\preprint{}

\title{\boldmath Pair production of charged Higgs 
bosons at future linear $e^{+}e^{-}$ colliders}

% Optional argument for running titles on pages
%\title[]{}



\author{Marco Battaglia}
\email[]{marco.battaglia@cern.ch}
%\thanks{}
\affiliation{CERN, Geneva, Switzerland}

\author{Arnaud Ferrari}
\email[]{ferrari@tsl.uu.se}
%\thanks{}
\affiliation{Department of Radiation Sciences, Uppsala University, Sweden}

\author{Ari Kiiskinen}
\email[]{ari.kiiskinen@cern.ch}
\author{Tuula M\"aki}
\email[]{tuula.maki@cern.ch}
%\thanks{}
\affiliation{Helsinki Institute of Physics, Helsinki, Finland}


\date{April 25, 2002 (Rev. Version)}

\begin{abstract}
In this paper, the pair production of charged Higgs bosons at possible 
future $e^{+}e^{-}$ linear colliders is studied. Multi-jet final states are 
considered and the combinatorial, hadronic and genuine $tbtb$ backgrounds 
are reduced thanks to a kinematical fit. TeV-class and multi-TeV linear 
colliders are likely to be sensitive to charged Higgs bosons with masses 
up to $\simeq$~350~GeV and 1.0~TeV respectively.
\end{abstract}
% \pacs{}

%\maketitle must follow title, authors, abstract and \pacs
\maketitle

\section{Introduction}

The exploration of the origin of mass and of the mechanism of electro-weak symmetry
breaking represents one of the main issues on the physics agenda of future colliders.
If the Higgs mechanism is indeed responsible for providing matter and force particles
with their masses, the detailed exploration of the Higgs sector will most likely 
require the combination of data from the {\sc Lhc} and lepton colliders. Today there 
are compelling indications that New Physics is required beyond the Standard Model (SM). 
In several SM extensions, a second complex scalar Higgs 
doublet is added yielding five physical Higgs bosons, two of which charged 
$H^{\pm}$, instead of just one. This is the case in the minimal supersymmetric SM 
extension (MSSM) and also in non-supersymmetric models (2HDM). The extra Higgs doublet 
introduces an additional fundamental parameter, describing the ratio of the 
vacuum-expectation values of the two doublets: $\tan \beta = v_2/v_1$. 

The negative results of charged Higgs boson searches in the {\sc Lep-2} and 
{\sc Tevatron} data and the indirect limits from the rate of the 
$b \rightarrow s \gamma$ process have set lower 
bounds on its  mass, $M_{H^{\pm}}$, but the bulk of the parameter 
space remains still unexplored. At the {\sc Lhc}, charged Higgs bosons can be observed 
but, according to present studies, only in the limited region of the parameters defined 
by $\tan \beta > 10 - 30$, depending on its mass and detectable decay modes.

In this paper, we discuss the potential of a high energy, high luminosity $e^+e^-$ linear
collider (LC) in the discovery of the charged Higgs boson and in the study of its 
properties. Since $M_{H^{\pm}}$ is not constrained in the models and can vary from the 
present 78.5~GeV {\sc Lep-2} limit~\cite{lep} up to, and beyond, 1~TeV, different 
collider energies need to be considered. 
We concentrate on a TeV-class LC able to deliver $e^+e^-$ collisions at 
$\sqrt{s}$ = 0.8~TeV, such as {\sc Tesla}~\cite{tesla} or an X-band LC~\cite{nlc}, and 
a multi-TeV LC operating at $\sqrt{s}$ = 3~TeV, such as {\sc Clic}~\cite{clic}, and 
discuss the experimental issues specific to the different LC parameters.

Two mass values have been chosen for the charged Higgs boson: 
$M_{H^{\pm}}$ = 300~GeV with moderate $\tan \beta$, representative of SUSY 
scenarios with a relatively light Higgs sector, and $M_{H^{\pm}}$ = 880~GeV with
$\tan \beta$ = 35, as point J from a set of SUSY benchmarks recently 
proposed~\cite{msugra}.

\section{Production cross section for charged Higgs bosons in 
$e^{+}e^{-}$ collisions}

In $e^+e^-$ collisions, charged Higgs bosons are pair produced via an 
intermediate photon or $Z^0$ boson. At tree level, the production cross section 
for $e^{+}e^{-} \rightarrow H^{+}H^{-}$ only depends on $M_{H^{\pm}}$ and on the 
centre-of-mass energy $\sqrt{s}$:
\[
\sigma = \frac{e^{4}}{48 \pi s}
\left( 1 - \frac{4m_{H}^{2}}{s} \right)^{3/2}
\left( 1 + \frac{2c'_{V}c_{V}}{1-m_{Z}^{2}/s} + 
\frac{{c'_{V}}^{2}(c_{V}^{2}+c_{A}^{2})}{(1-m_{Z}^{2}/s)^{2}} \right)
\]

However, in calculating the effective cross sections at high energy 
$e^{+}e^{-}$ collisions, initial state radiation (ISR) and beam-beam 
effects must be taken into account. 
Because of the very small transverse dimensions of the beams at the interaction point, 
the electrons and positrons undergo significant radiation, in the field of the 
incoming beam, before collision (beamstrahlung). The average energy loss is of 
the order of 5\% for $\sqrt{s}$=0.8~TeV at {\sc Tesla} but reaches 31\% for 
$\sqrt{s}$=3~TeV at {\sc Clic}. 

\begin{figure}[h!]   
\begin{center} 
\includegraphics[height=7.0cm]{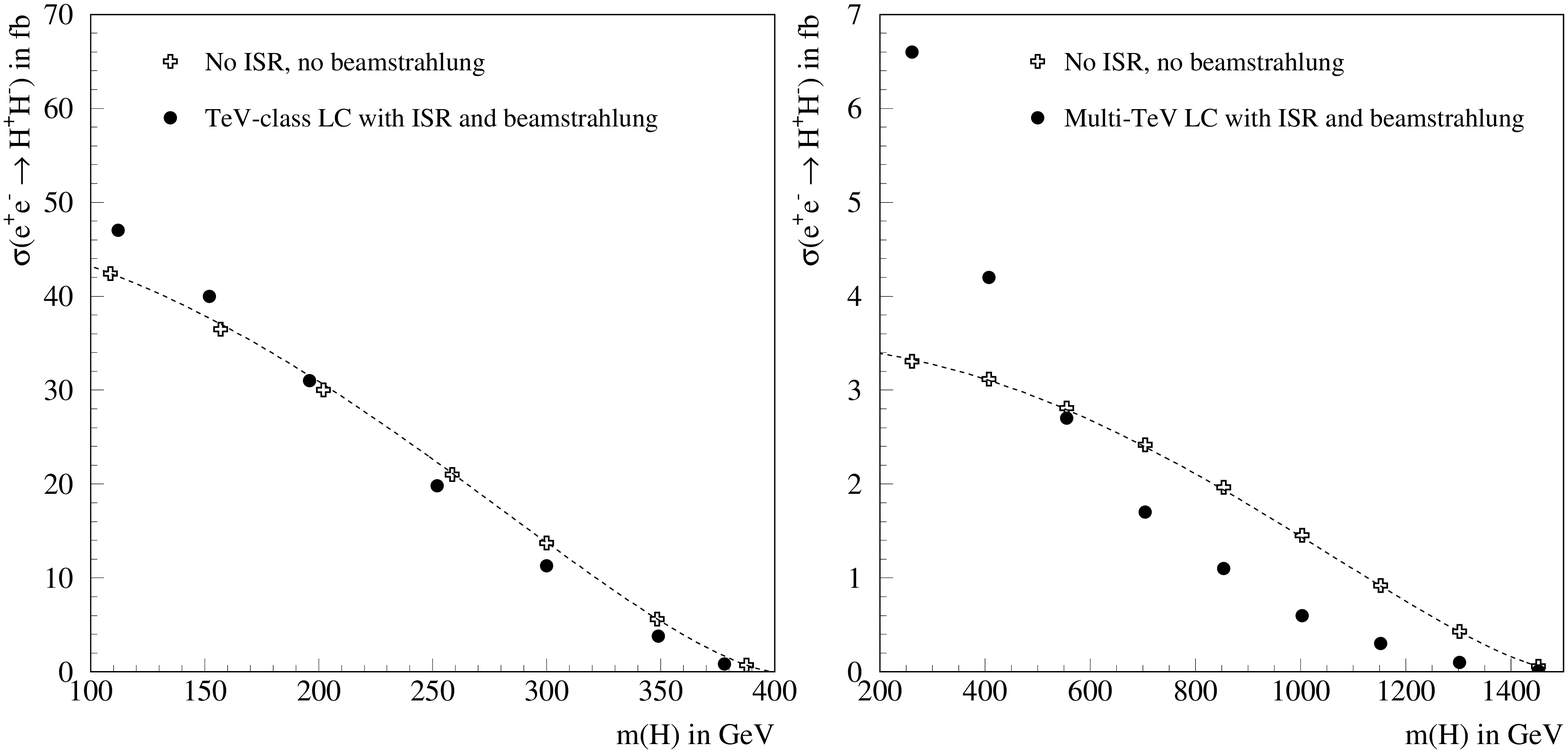}
\caption[]
{Cross section for $e^{+}e^{-} \rightarrow H^{+}H^{-}$ at 
$\sqrt{s}=800~\mbox{GeV}$ (left) and at $\sqrt{s}=3~\mbox{TeV}$ 
(right). Open crosses and dashed lines show the cross section 
at tree level, while full circles show the effective cross 
section after including ISR and beamstrahlung.}
\label{fig:crosshh}
\end{center}
\end{figure}

When these effects are taken into account, the tree level cross section is folded 
with the luminosity spectrum. For the low values of $M_{H^{\pm}}$ the effective 
cross section is enhanced compared to the tree level cross section, while close to the 
kinematical threshold only a smaller fraction of the energy spectrum is available and 
the effective cross section is reduced as exemplified in Figure~\ref{fig:crosshh}.
Corrections to the $H^+H^-$ production cross section have been computed at one loop and 
found to become sizeable in some region of the parameter space~\cite{hhcorr}. 
A precise determination of the cross section can thus provide additional information on
model parameters.

\begin{figure}[h!]   
\begin{center} 
\includegraphics[height=9.0cm,width=11.0cm]{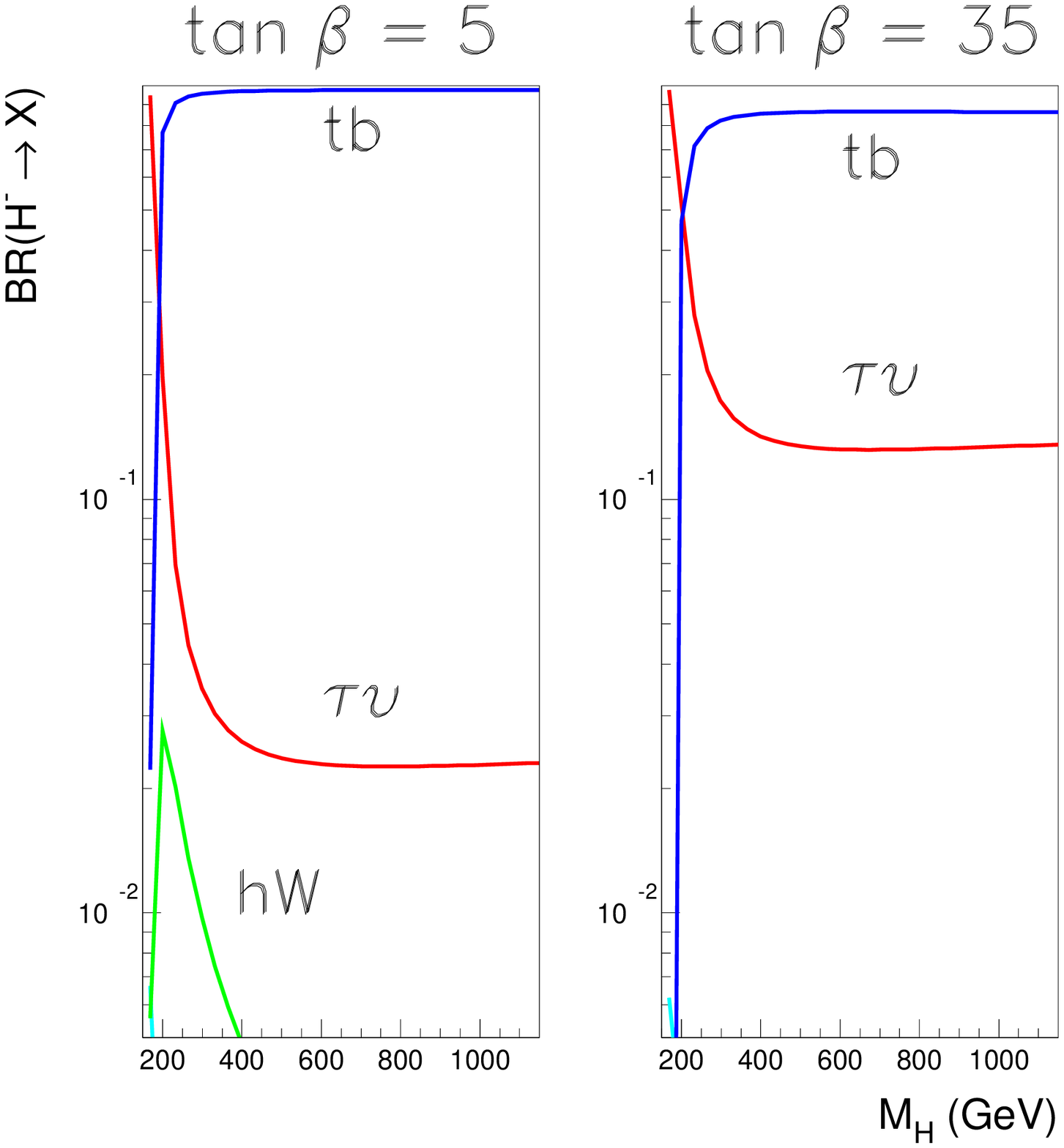}
\caption[]
{$H^{\pm}$ branching fractions for the dominant decay modes as a function of 
$M_{H^{\pm}}$ for two values of $\tan \beta$.}
\label{hdecay}
\end{center}
\end{figure}

When considering only decays to SM particles, the dominant $H^-$ decay modes are
those involving the heaviest quark or lepton pairs accessible, i.e. $\bar t b$ and 
$\tau^- \bar \nu_{\tau}${\footnote{In the following the notation of a particle implies 
also its anti-particle}}, as shown in  Figure~\ref{hdecay} for two values of 
$\tan \beta$. The dependence of the dominant branching fractions on $\tan \beta$ can be 
exploited for determining this fundamental parameter of the theory. 
In this study we consider the fully hadronic
$e^{+}e^{-} \rightarrow H^{+}H^{-} \rightarrow (t\bar{b})(\bar{t}b)$ channel, 
which leads to two $W$ bosons and four $b$ quarks in the final state and provides with 
a direct reconstruction of $M_{H^{\pm}}$.

\section{Event Reconstruction}

The {\sc Simdet} parametrised detector simulation~\cite{ref-simdet} has been 
used to account for the anticipated experimental resolutions of a LC detector.
The $b$-tagging performance, which is of major importance in this analysis, has been 
included, assuming an efficiency $\epsilon_b$ = 0.90.
The characteristic signal final state, including four $b$ jets and the intermediate
$W$ and $t$ mass constraints, guarantees an efficient rejection of most multi-fermion 
background processes. The cross section of the irreducible 
$t \bar b \bar t b$ background has been  estimated using the {\sc Comphep} 
program~\cite{comphep} at 0.8~TeV and 3~TeV and found to be 5.5~fb and 1.5~fb 
respectively. 

Hadronic jets have been reconstructed using the CAMJET clustering algorithm, for the 
0.8~TeV analysis, and the Lund algorithm for 3~TeV, tuning the cut-off to maximise the 
number of eight jet events for the signal.
In order to distinguish the production of charged Higgs bosons from 
the associated background processes and to accurately measure its mass, 
it is important to obtain a clean mass distribution of the multi-jet 
final states. Therefore only hadronic $W$ boson decays have been considered
by removing events with significant transverse missing energy or 
tagged leptons. 
 
The mass reconstruction has been performed as follows.
In a first step, the presence of two $W$ bosons decaying hadronically 
has been tested. The assignment of the four non-$b$ tagged jets to their 
correct $WW$ pair has been done by choosing the combination which minimises 
the difference between the di-jet masses and the $W$ mass. In a second step, 
each $t$ quark has been reconstructed from a $W$ candidate paired with one of the 
four $b$ tagged jets, taking the combination giving the invariant mass closest to 
the top mass. Each of the charged Higgs bosons has then been
reconstructed from a $t$ candidate paired with one of the two remaining 
$b$ tagged jets and selecting the configuration minimising the mass difference of 
the two $tb$ pairs. Events with no jet combination compatible with the $W$ or $t$ 
masses within the observed resolution have been discarded.
An event kinematical fit has been finally applied to improve the mass 
resolution. The fit uses energy and momentum conservation, the $W$ and $t$ mass 
constraints and imposes $H$ boson equal mass.

\subsection{$M_H$ = 300~GeV at a TeV-class LC}

\begin{figure}[h!] 
\includegraphics[height=7.5cm]{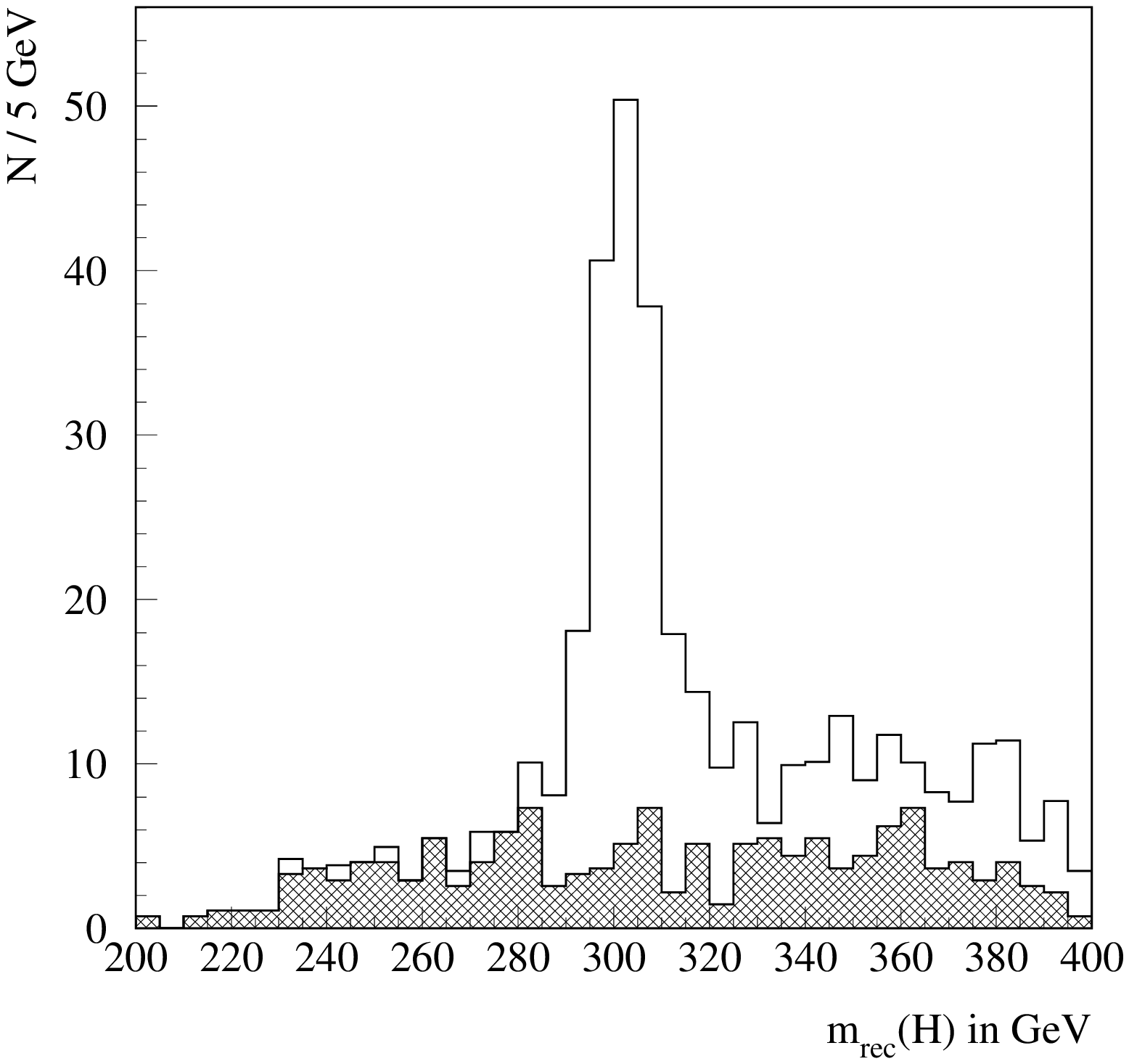}
\caption{Fitted charged Higgs boson mass for $H^+ H^- \rightarrow 
(t\bar{b})(\bar{t}b)$ with $m_{H}$ = 300~GeV. The histogram is normalised 
to an integrated luminosity of 1~$\mbox{ab}^{-1}$, with a 100\% branching 
ratio in the analysed decay mode. The contribution of the $tbtb$ 
background events is shown by the dark histograms.}
\label{tesla_mass}
\end{figure}

At $\sqrt{s}$ = 0.8~TeV, the reconstructed mass peak has a width of 6.2~GeV with a 
natural $H^{\pm}$ width of 4.3~GeV, for $M_{H^{\pm}}$ = 300~GeV. This corresponds to a 
mass resolution of 1.5\%. The signal reconstruction efficiency has been found to be 
0.022, corresponding to 149 signal with 24 background events, in a signal mass window of
$\pm2.5\sigma$ around the mass peak. The good mass resolution achieved with the 
kinematical fit gives a signal significance $S/\sqrt{B}$=30 
(see Figure~\ref{tesla_mass}). The product 
$\sigma(e^+e^- \rightarrow H^+H^-) \times {\mathrm{BR}}(H^- \rightarrow \bar t b)$ 
can be measured to an accuracy of 8.8\%. 

The maximum reach in mass for a $5~\sigma$ discovery has been studied by analysing
signal samples generated at different Higgs mass. The largest observable Higgs mass 
corresponds to about 350~GeV, corresponding to the detection of 47 signal events, with 
the same number from background for 1~ab$^{-1}$ of data at $\sqrt{s}$ = 0.8~TeV.

At moderate $\tan \beta$ values, as well as in scenarios where the couplings to fermions 
are suppressed, the $H^{-} \rightarrow W^-h^0$ decay, leading to the same 
$W^- b \bar b$ decay final state, may become sizeable. In order to study  this process, 
a similar mass reconstruction procedure can be applied except for the top quark 
constraint being replaced  by the $h^0$ one. By comparing the $\chi^2$ of the fit for 
the two hypotheses, the $H^{-} \rightarrow \bar t b$ and $H^{-} \rightarrow W^-h^0$
decays can be efficiently distinguished.

\subsection{$M_H$ = 880~GeV at a multi-TeV LC}

At a multi-TeV collider the decay reconstruction is complicated by the larger 
beamstrahlung, as mentioned above, and  by the $\gamma \gamma \rightarrow 
{\mathrm{hadrons}}$ background~\cite{bkg} overlayed to each $e^+e^-$ collision.

The reconstruction procedure for the charged Higgs boson follows closely that described
above, but special care has been taken to make the result robust in presence of the 
$\gamma \gamma$ background, while the broader luminosity spectrum affects the kinematic 
fit, as discussed below.
The additional hadrons generated by these $\gamma \gamma$ collisions, 
mostly affecting the forward regions, 
may either be merged into the jets coming from the $H^{\pm}$ decay or result in extra 
jets being reconstructed. It is therefore important to minimise the impact from this 
background source on the event reconstruction. In this analysis only the four leading 
non-$b$ jets have been considered, together with the $b$-tagged jets.
In addition the resulting invariant masses of the intermediate 
states are biased towards higher values due to the additional hadron contribution. 
This needs to be taken into account by calibrating the mass response with simulation 
including the $\gamma \gamma$ background.

\begin{figure}[h!]   
\begin{center} 
  \includegraphics[height=7.5cm]{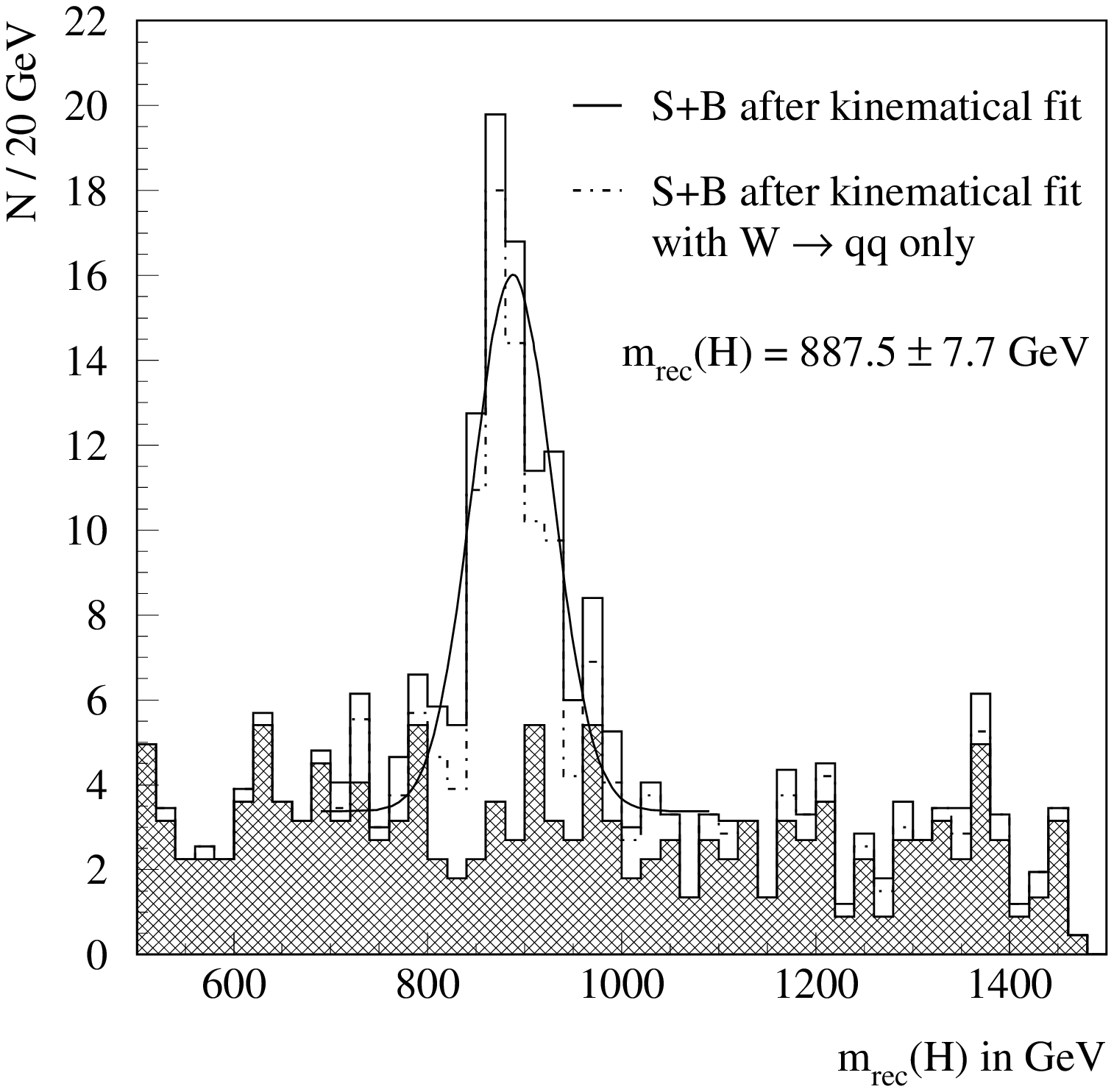}
\caption[]
{Reconstructed $H^{\pm} \rightarrow tb$ invariant mass after the kinematical fit for 
an integrated luminosity of 3~ab$^{-1}$. The $tbtb$ background is shown by the 
hatched distribution.} 
\label{higgs-recplots}
\end{center}
\end{figure}

Due to the significant loss of energy of the colliding $e^+$ and $e^-$, energy and 
momentum conservation constraints cannot be applied on the reconstructed system for 
the nominal $\sqrt{s}$. Instead the kinematical fit allows for an extra 
particle to be radiated but imposes its transverse momentum to be zero. The kinematic 
fit allows to improve, by a factor of two, the resolution on the mass reconstruction for
$H$ candidates. Furthermore, the use of the 
kinematical fit, allows to correct for the effect of the $\gamma\gamma$ background 
resulting in a mass peak of the charged Higgs boson at the position of the generated 
mass (see Figure~\ref{higgs-recplots}).

The signal event rate has been estimated for 
$M_{H}=880~\mbox{GeV}$, first disentangling the $\gamma\gamma$ background 
and then assuming that the detector integrates the background
from 15 bunch crossings (BX). 
In the first case with an integrated luminosity is 3~ab$^{-1}$ and assuming 
BR($H \rightarrow tb$) = 1.0, the r.m.s. of the mass peak, after the kinematical fit, 
is 33~GeV with a $H^{\pm}$ natural width of 21~GeV.
47 $e^{+}e^{-} \rightarrow H^{+}H^{-} \rightarrow (t\bar{b})(\bar{t}b)$ signal
events are expected to be reconstructed within the $\pm 2.5\sigma$ signal mass region, 
corresponding to a signal efficiency of $\simeq~0.02$. 

When the hadronic background, integrated over 15~BX is included, the peaks of the 
intermediate $W$ and $t$ states are shifted towards higher masses and get
significantly wider. To account for this effect, the mass windows for the reconstructed 
$W$ boson and $t$ quark need to be shifted and made 50\% wider. The fitted signal 
width becomes 42~GeV and there are 51 $e^{+}e^{-} 
\rightarrow H^{+}H^{-} \rightarrow (t\bar{b})(\bar{t}b)$ signal events with a
reconstructed mass in the signal window, corresponding to an efficiency
comparable to the previous case. The $tbtb$ background processes contribute 24 events 
for 3~ab$^{-1}$. By scaling these results, the estimated maximum mass 
reach for a $5~\sigma$ discovery with 3~ab$^{-1}$ at 3~TeV is $\simeq$1.0~TeV.

Beyond discovery the properties of the charged Higgs boson needs to be studied. These
include the mass, the production rate and the decay branching fractions. The mass can 
be measured to an accuracy of 1\% for 15 integrated BX and the product 
$\sigma(e^+e^- \rightarrow H^+H^-) \times {\mathrm{BR}}(H^- \rightarrow \bar t b)$ can be
measured to an accuracy of $\simeq$~20\%. 

\section{Conclusions}

The full reconstruction of the charged Higgs boson $H^+H^- \rightarrow
t\bar{b} \bar{t}b$ multi-jet final states has been studied for two scenarios:
$M_{H^{\pm}}$ = 300~GeV at $\sqrt{s}$ = 800~GeV and $M_{H^{\pm}}$ = 880~GeV with 
$\sqrt{s}$ = 3~TeV. The analyses are based on the reconstruction of hadronic events 
with eight jets, four of which $b$-tagged. A kinematical fit taking into account the
$W$ and $t$ intermediate mass constraints and energy-momentum conservation has been 
applied to reject the multi-fermion background and optimise the mass resolution for 
signal events. The Higgs mass can be reconstructed with an accuracy of $\simeq~1\%$
and the product $\sigma(e^+e^- \rightarrow H^+H^-) \times 
{\mathrm{BR}}(H^- \rightarrow \bar t b)$ can be measured to an accuracy of 8.8\% for 
$M_{H^{\pm}}$ = 300~GeV at 0.8~TeV and 15\% for $M_{H^{\pm}}$ = 880~GeV at 3~TeV.

For further studies, a more efficient event selection needs to be envisaged. 
An interesting possibility is to tag the 
$e^+e^- \rightarrow H^+H^-$ event by reconstructing only one $H \rightarrow tb$ decay. 
This allows to increase the event selection efficiency and to study the $H$ decay in 
the hemisphere opposite to the tagged one in an unbiased way.

\begin{acknowledgments}
We are grateful to E.~Boos for his help with the {\sc Comphep} package. 

The research activity of A.F. has been supported by a Marie Curie Fellowship of the 
European Community programme {\sl Improving Human Research Potential and the 
Socio-economic Knowledge Base} under contract number HPMF-CT-2000-00865.
\end{acknowledgments}
%\bibliography{hh-snowmass.bib}

\end{document}